\documentstyle{article}
\begin{document}
\title{On the circular polarization of pulsar radiation}
\author{ Y.E.~Lyubarskii and S.A.~Petrova\\
\small Institute of Radio Astronomy, 4 Chervonopraporna St.,
Kharkov,\\  \small 310002 Ukraine}
\date{ }

\maketitle
\centerline{\bf Abstract}
We consider the polarization behaviour of radio waves propagating through
an ultrarelativistic highly magnetized electron-positron plasma in a
pulsar magnetosphere. The rotation of magnetosphere gives rise to the wave mode
coupling in the polarization-limiting region.
The process is shown to cause considerable circular
polarization in the linearly polarized normal waves. Thus, the circular
polarization observed for a number of pulsars, despite the
linear polarization of the emitted normal
waves, can be attributed to the
limiting-polarization effect.

\section{Introduction}
Pulsar radiation is known to have a high linear polarization. Although the
circular polarization of the average pulse profiles is typically less than the 
linear one, for the most of pulsars the circular polarization is also 
significant (for recent review, see Han {\it et al.} 1998 and references 
therein). The circular
polarization tends to peak near the centre of the profile. In many pulsars
the circular polarization reverses its sense near the pulse centre, while
in the others the sense remains constant throughout the pulse 
(Radhakrishnan \& Rankin 1990). The observed circular polarization of
pulsar radiation is commonly attributed either to the propagation effects or
to the emission mechanism.

The magnetic field pattern of a pulsar is generally considered to be that
of a rotating dipole. The magnetosphere contains an electron--positron
plasma outflowing along the open magnetic field lines with Lorentz-factors
$\gamma\sim 100$.
So pulsar radiation, believed to originate in the open field line tube,
propagates through the highly magnetized relativistic plasma. The propagation
of radio waves in a relativistically moving medium was analyzed by Lee \&
Lerche (1975), Harding \& Tademaru (1981)
However, they considered rather abstract
model of a medium, with the spatial dispersion being ignored. A qualitative
consideration of wave polarization behaviour in application to a more
realistic pulsar model allowed to attribute a number of characteristic
features of the observed polarization pattern to the propagation effects
inside the magnetosphere (Cheng \& Ruderman 1979, Stinebring 1982, Barnard
1986, von Hoensbroech {\it et al.} 1998).

Given that the plasma is embedded in an infinitely strong magnetic field,
the normal wave modes are linearly polarized. One of them, namely the
ordinary mode, is polarized within the same plane as the wave vector, ${\bf k}$,
and the magnetic field, ${\bf b}$; the other one, called the extraordinary
mode, is polarized perpendicularly to the this plane.
In the emission region, the scale length for beats between the wave modes
is much less than the scale length for change in the plasma parameters.
So in the vicinity of the emission origin the normal waves propagate
independently, their polarization plane being adjusted to the local
orientation of the magnetic field. Along the trajectory, the angle $\theta$
between the wave vector and the magnetic field increases due to
magnetic line curvature and the plasma density decreases due to open field
line tube widening, the scale length for beats increasing. Eventually the
plasma density decreases enough and the medium does not influence the waves;
their polarization remains now fixed. Thus, the emergent polarization is set
up by the processes in the so--called polarization--limiting region, where
the geometrical optics approximation is violated (Budden 1952). The
polarization--limiting radius, $r_p$, can be found from the relation:
$$
\frac{\omega}{c}\Delta n(r_p)s(r_p)\sim 1,          \eqno (1.1)$$
where $\Delta n$ is the difference between the refractive indices of the wave
modes considered, $\omega$ the wave frequency, $s$ the scale length for
change in the plasma parameters, $s\sim r_p$. In the highly magnetized
relativistic plasma of number density $N$ and Lorentz-factor $\gamma$,
the refractive indices for the ordinary and extraordinary modes are known
to be, respectively (Melrose \& Stoneham 1977):
$$
n_o=1-\frac{\omega_p^2\sin^2\theta}{2\omega^2\gamma^3(1-\beta\cos\theta)^2},
\quad\qquad n_e=1, \eqno (1.2)$$
with $\omega_p\equiv\sqrt\frac{\textstyle 4\pi Ne^2}{\textstyle m}$ being the plasma frequency,
$\beta$ the plasma velocity in units of $c$.

In application to pulsars, the polarization-limiting radius was estimated by
Cheng \& Ruderman (1979), Stinebring (1982) and Barnard (1986). Note that
if a normal wave were emitted in the field line plane of a non-rotating
dipolar magnetic field, the polarization plane would not vary along the
trajectory at all; the normal modes then would propagate independently
preserving their initial polarization states. As the wave propagates in the
rotating magnetosphere, the wave vector acquires a tilt to the ambient
magnetic line planes, so that the ${\bf k}\times{\bf b}$--plane turns along
the trajectory. Another consequence of rotation, namely the magnetic line
sweepback, also leads to rotation of the ${\bf k}\times{\bf b}$--plane, however,
this effect is less significant. As the scale length for the wave mode
beats increases sufficiently, the polarization of a normal wave has no time
to follow the variations of the ambient magnetic field and the wave mode
coupling holds. Cheng \& Ruderman (1979) pointed out that the propagation
of the linearly polarized normal waves in the region where the geometrical
optics fails, results in elliptical polarization of the emergent radiation.
The effect is likely to account for the circular polarization observed in
pulsars that exhibit predominantly one sense of circular polarization
across the pulse (Radhakrishnan \& Rankin 1990).

The influence of pulsar
rotation on the observed polarization was also considered by Blaskiewicz
{\it et al.} (1991). These authors investigated the contribution of rotation
to the velocity of the particles emitting curvature radiation, with the
propagation effects being neglected. However, it is the wave propagation
through the magnetosphere that sets up the characteristics of outgoing
radiation.
Indeed, whatever the emission mechanism, only the waves
corresponding to the modes allowed by the magnetospheric plasma can propagate
and ultimately escape from pulsars.

Provided that the plasma with the various distribution functions for 
electrons and positrons is embedded in the finite magnetic field, the normal 
waves propagating at the small angle to the field should be circularly
polarized. Proceeding from this, Cheng \& Ruderman (1979), von 
Hoensbroech {it et al.} (1998) proposed to explain the observed circular 
polarization of pulsar radiation by the dispersive properties of the
magnetospheric plasma. However, well within the pulsar magnetosphere the
magnetic field strength is so high that the critical angle for the
circularly polarized normal modes appears to be too small.


The present paper deals with a more detailed consideration of wave
polarization behaviour in the polarization-limiting region. In Sect.\ 2 we
treat the equations describing the waves which propagate  through an
ultrarelativistic highly magnetized plasma rotating together with the magnetic
field. In Sect.\ 3 these equations are applied to the pulsar magnetosphere.
The wave mode coupling in the polarization-limiting region is found to
produce a significant circular polarization in the initially linear
normal waves. The results are summarized in Sect.\ 4.

\section{Basic equations}
Consider an ultrarelativistic electron-positron plasma rotating together
with an infinitely strong inhomogeneous magnetic field. For the waves of
frequency $\omega$ the wave fields ${\bf E}$, ${\bf B}$ are described by
Maxwell's equations:
$$
\nabla\times {\bf B}=-\frac{i\omega}{c}{\bf E}+\frac{4\pi}{c}{\bf j}_1,$$
$$
\nabla\times {\bf E}=\frac{i\omega}{c}{\bf B},
\eqno (2.1)$$
$$
-i\omega e(n_1^+-n_1^-)+{\rm div} {\bf j}_1=0.
$$
Here ${\bf j}_1$ is the linearized current density caused by the waves,
$$
{\bf j}_1\equiv e[{\bf v}_0(n_1^+-n_1^-)+n_0({\bf v}_1^+-{\bf v}_1^-)],
\eqno (2.2) $$
with ${\bf v}_0$, $n_0$ being the unperturbed particle velocity and
number density, respectively, ${\bf v}_1^\pm$, $n_1^\pm$ the small
perturbations of these quantities for electrons and positrons. Although the
distribution functions for electrons and positrons are generally believed
to be slightly different, the difference is insignificant in our case.
Therefore we assume that ${\bf v}_0$ and $n_0$ are equal for both the
particle species.

In the infinitely strong magnetic field, the particle motion can be treated
in terms of the mechanical bead-on-a-wire model. The particle moves along
the magnetic field line which rotates at the angular velocity ${\bf \Omega}$.
The particle velocity in the laboratory frame may be written as
$$
{\bf v}={\bf \Omega}\times {\bf r}+v_b{\bf b}, \eqno (2.3) $$
where ${\bf b}$ is the unit vector along the magnetic field, $v_b$ the
velocity along the field line. Substituting Eq.\ (2.3) into the
Lagrangian
$$
L=-mc^2\sqrt{1-\frac{v^2}{c^2}}+\frac{e}{c}({\bf A}\cdot {\bf v})-e\varphi ,
$$
with ${\bf A}$ and $\varphi$  being, respectively, the vector and scalar
potentials of the wave, one can get the equation of motion. To the first
order in $\frac{\textstyle \vert{\bf \Omega}\times {\bf r}\vert}{\textstyle c}$ we obtain:
$$
m\gamma^3\frac{dv_b^\pm}{dt}=\pm e\left[{\bf E}
+\frac{({\bf \Omega}\times{\bf r})
\times{\bf B}}{c}\right]\cdot {\bf b}. \eqno (2.4)$$
Here the curvature radius of the magnetic field lines was assumed to be much
higher than the wavelength; indeed, this is a good approximation in our case.
The equation of motion  (2.4) implies that the inertial forces introduced by
the rotation are small as $O\left(\frac{\textstyle \vert {\bf \Omega}\times{\bf r}\vert^2}
{\textstyle c^2}\right)$. Note that the left-hand side of Eq.\ (2.4) contains the total
derivative, $\frac{\textstyle d}{\textstyle dt}\equiv -i\omega +{\bf v}_0\cdot \nabla $.

Equations (2.1), (2.2) and (2.4) yield the self-consistent description of the
wave
fields and the plasma particle motion in these fields. The refractive indices
for the waves in the polarization-limiting region equal unity to within
$\frac{\textstyle c}{\textstyle \omega r}\ll 1$. Therefore
the waves propagate almost straight-line and one can choose the
three-dimensional Cartesian system with the z-axis aligned with the wave
vector. Then all the perturbed quantities should depend only on the
z-coordinate. On the basis of Eqs.\ (2.1) and (2.4) one can write the
component equations:
$$
\frac{d^2E_x}{dz^2}+\frac{\omega^2}{c^2}E_x+\frac{4\pi i\omega}{c^2}j_{1x}=0,$$
$$
\frac{d^2E_y}{dz^2}+\frac{\omega^2}{c^2}E_y+\frac{4\pi i\omega}{c^2}j_{1y}=0,$$
$$
\frac{\omega^2}{c^2}E_z+\frac{4\pi i\omega}{c^2}j_{1z}=0, \eqno (2.5)$$
$$
-i\omega e(n_1^+-n_1^-)+\frac{dj_{1z}}{dz},$$
$$
m\gamma^3\left(-i\omega v_1^\pm+v_{0z}^\pm\frac{dv_1^\pm}{dz}\right)$$
$$=\pm e\left[{\bf E}-
\frac{i({\bf \Omega}\times{\bf r})\times(\nabla\times {\bf E})}{\omega}\right]
\cdot {\bf b}. $$

We consider the waves propagating along the z-axis in positive direction.
Since the refractive indices of the waves considered are very close to
unity, the spatial dependence should be close to $\exp\left(i\frac{\omega}{c}
z\right)$.
Then one can present the field components in the form:
$$
E_\mu =a_\mu\exp{\left(i\frac{\omega}{c}z\right)}, \eqno (2.6)$$
with the amplitudes $a_\mu $ varying slowly:
$$
\frac{da_\mu }{dz}\ll\frac{a_\mu \omega}{c}. \eqno (2.7)$$
Of course, the rest of the perturbed quantities can be presented similarly.
The scale length for change in the medium parameters, namely in ${\bf b}$,
${\bf v}_0$ and $n_0$, also  exceeds the wavelength essentially. Then the set
of equations (2.5) is reduced to the form:
$$
\begin{array}{l}
\frac{da_x}{dz}=-iR[(b_x+q_y)^2a_x+(b_x+q_y)(b_y-q_x)a_y],\\
\frac{da_y}{dz}=-iR[(b_x+q_y)(b_y-q_x)a_x+(b_y-q_x)^2a_y],
\end{array}
\eqno (2.8)$$
where
$$
{\bf q}\equiv \frac{{\bf b}\times ({\bf \Omega}\times {\bf r})}{c},
\qquad R\equiv\frac{\omega_p^2}{2\omega c\gamma^3(1-\beta_z)^2},
$$
with $\omega_p\equiv\sqrt{\frac{\textstyle 8\pi n_0e^2}{\textstyle m}}$ being the plasma frequency,
$\beta_z$ the z--component of the plasma velocity  ${\bf v}_0$ in units of $c$.
Above we took into account that the waves considered are quasi--transverse
ones, $1-n\ll 1$, and, correspondingly, the z--component of the wave electric
field is small, $a_z\ll a_x$, $a_y$. With $n$ given by Eq.\ (1.2), this is
surely valid at the condition (1.1).

Neglecting the rotation and the z-dependence of unperturbed plasma
parameters we come to the homogeneous problem. Then the solutions are easy to
be found in the form:
$$
a_\mu(z)\propto \exp \left(-i\frac{\omega}{c}(1-n)z\right). \eqno (2.9)$$
Setting the determinant of the system equal to zero immediately yields the
customary refractive indices (1.2) for the ordinary and
extraordinary modes in the homogeneous highly magnetized ultrarelativistic
plasma.

\section{Polarization transfer in the rotating magnetosphere}
\subsection{The main features of wave propagation}
First we examine the set of equations (2.8) to outline the characteristic
features of wave polarization behaviour along the trajectory in pulsar
magnetosphere. As long as the plasma density is sufficiently high, so that
$Rz\gg 1$, geometrical optics approximation is valid. Then the wave field
amplitudes can be presented in the form:
$$
a_\mu =a_\mu^{(0)}\exp [G(z)], \eqno (3.1)$$
with $G(z)$ being as large as $Rz$. Substituting Eq.\ (3.1) into Eq.\ (2.8)
and setting the determinant of the system equal to zero one can find, to the
first order in $(Rz)^{-1}$,
$$
\left(\frac{dG}{dz}\right)_o=-iR[(b_x+q_y)^2+(b_y-q_x)^2],$$
$$\left(\frac{dG}{dz}\right)_e=0, \eqno (3.2)$$
with the subscripts $o$ and $e$ referring to the ordinary and extraordinary
modes, respectively. Note that the terms $q_x$ and $q_y$ are introduced by
the magnetosphere rotation, so that in the corotating frame $q_x=q_y\equiv 0$.
Comparison of Eqs.\ (3.1), (3.2) with Eqs.\ (1.2), (2.9) corresponding to
the homogeneous plasma then leads to a well--known result: the wave vector
in a weakly inhomogeneous medium equals, in each point of the trajectory,
to that in the homogeneous medium with the same parameters, so that
$a_\mu\propto\exp \left[-i\frac{\omega}{c}\int^z(1-n(z))dz\right]$. With
Eq.\ (3.2) we obtain:
$$
\left(\frac{a_x^{(0)}}{a_y^{(0)}}\right)_o=\frac{b_x+q_y}{b_y-q_x},\qquad
\left(\frac{a_x^{(0)}}{a_y^{(0)}}\right)_e=-\frac{b_y-q_x}{b_x+q_y}.
\eqno (3.3)$$
Hence, in the corotating frame the polarization of normal waves follows the
slow variations of the ambient magnetic field.

Since the plasma density decreases along the trajectory, $N\propto z^{-3}$,
geometrical optics ultimately fails. In the opposite limit, $Rz\ll 1$, wave
propagation is no longer influenced by the plasma and, obviously, wave
polarization remains fixed (cf. Eq.\ (2.8)). Thus, the emergent polarization
is formed in the so--called polarization--limiting region, where $Rz\sim 1$.
In this region, the scale length for wave mode beats becomes so large that
the polarization of a normal wave has no time to follow the variation of the
ambient magnetic field, so that the wave mode coupling holds.

As follows straightly from Eq.\ (2.8), if the magnetosphere were
non--rotating one, the normal waves emitted in the plane containing the
field lines would propagate independently throughout the trajectory, the
polarization plane remaining constant. Now we are interested in  the wave mode coupling
introduced by the slow magnetosphere rotation. The rotation causes the wave
vector tilt of $\sim z/r_L$ to the ambient field line planes along the
trajectory; here $r_L\equiv\frac{\textstyle c}{\textstyle \Omega}$
is the light cylinder radius. In addition, the dipolar magnetic field
structure should be distorted. Indeed, the poloidal current flow,
$j\sim\frac{\Omega B_p}{\textstyle 2\pi}$,
with $B_p$ being the poloidal magnetic field strength (Goldreich \&
Julian 1969), gives rise to
the azimuthal component of the magnetic field, $B_\varphi$:
(rot${\bf B})_p\sim\frac{\textstyle B_\varphi}{\textstyle d}\sim\frac{\textstyle
4\pi}{\textstyle c}j$, where $d$ is the open field line tube diameter,
$d=\frac{\textstyle z^{3/2}}{\textstyle r_L^{1/2}}$. The field line twist
then can be estimated as $\frac{\textstyle B_\varphi}{\textstyle B_p}\sim \left(
\frac{\textstyle z}{\textstyle r_L}\right)^{3/2}$. Thus, the sweepback of
magnetic lines appears to be less significant than the rotation effect.
Therefore below we consider the rotation of purely dipolar magnetic field.

Let $\delta$ be the opening half-angle of the emission cone. Consider
the wave emitted along the z-axis at the angle $\delta$ to the
magnetic axis, the latter lying in the xz-plane at the moment of emission.
 Then the direction cosines of the non-rotating dipolar magnetic
field far from the emission point are
$$
b_{x0}=\delta/2,\qquad
b_{y0}=0, \qquad
b_{z0}=1-b_{x0}^2/2.
\eqno(3.4)$$
Now let the magnetic field rotate at the angular velocity ${\bf \Omega}$,
the rotational axis making the angles $\alpha$ and $\xi$ with the magnetic
axis and the wave vector, respectively. The open field line tube is believed
to be narrow, so that $\vert \alpha -\xi \vert\leq\delta\ll 1$. To the first
order in $\Omega t$, the direction cosines of the rotating field are then
given by
$$
b_x=b_{x0}+\Omega t\sin\xi\frac{\sqrt{\delta^2-(\alpha -\xi)^2}}{\delta},$$
$$
b_y=\pm\Omega t\sin\xi\frac{\alpha -\xi}{\delta}, \eqno (3.5)$$
$$
b_z=1-\frac{b_{x0}^2+b_{y0}^2}{2} ,$$
with $t\equiv z/c$. The sign of the component $b_y$
is that of the production ${\bf b}_d\cdot({\bf k}\times
{\bf \Omega})$; here ${\bf b}_d$ is aligned with the magnetic axis at the
moment of emission.
Using Eq.\ (3.5) and
taking into account that the waves are emitted close to the magnetic axis,
$x$, $y\ll z$, one can obtain the components of the vector
${\bf q}\equiv {\bf b}\times ({\bf \Omega}\times {\bf r})$:
$$
q_x=\mp\frac{\Omega z}{c}\sin\xi\frac{\alpha -\xi}{\delta},$$ $$
q_y=\frac{\Omega z}{c}\sin\xi\frac{\sqrt{\delta^2 -(\alpha -\xi)^2}}{\delta}
.\eqno (3.6) $$
The sign of $q_x$ is opposite to the sign of
${\bf b}_d\cdot({\bf k}\times{\Omega})$.

\subsection{Wave mode coupling in case $\frac{\textstyle z_p}{\textstyle r_L
\delta}\ll 1$}
First we assume that $\frac{\textstyle z_p}{\textstyle r_L\delta}\ll 1$. Then
taking into account that $\omega_p\propto N\propto z^{-3}$ and substituting
Eqs.(3.5), (3.6), one can reduce Eq.(2.8) to the form
$$
\frac{da_x}{du}-iua_x=i\frac{z_p}{r_L\delta}\eta (b_1ua_x\pm b_2a_y),$$
$$
\frac{da_y}{du}=\pm i\frac{z_p}{r_L\delta}\eta b_2a_x , \eqno (3.7)$$
where $u\equiv z_p/z$, with $z_p$ referring to the polarization--limiting
radius, $\eta\equiv 2\sin\xi$, $b_1\equiv\frac{\textstyle \sqrt{\delta^2
-(\alpha -\xi)^2}}{\textstyle \delta}$, $b_2\equiv\frac{\textstyle \alpha -
\xi}{\textstyle \delta} $. Above we used the definition of the
polarization-limiting radius (Eqs.\ (1.1), (1.2)) implying that
$$
\frac{\textstyle 8\omega_p^2(z_p)z_p}{\textstyle \omega c\gamma^3\delta^2}=1.
$$
The wave mode coupling caused by the magnetosphere rotation is described by
the right--hand sides in Eq.\ (3.7), which are small as $\frac{\textstyle
z_p}{\textstyle r_L\delta}$. It is easy to find the solutions of
Eq.\ (3.7) to the leading order in $\frac{\textstyle z_p}{\textstyle
r_L\delta}$. If, say, the ordinary mode is emitted, that is at the emission
origin $a_x=C_x$, $a_y=0$, the wave polarization can be found to vary
along the trajectory as follows:
$$
a_x=C_x\exp (iu^2/2),$$
$$
a_y=\pm i\frac{z_p}{r_L\delta}\eta b_2C_x\int^u\exp (iu^{\prime 2}/2)
du^\prime. \eqno (3.8) $$
Setting $z\to \infty$ ($u\to 0 $) yields the limiting polarization of the
emitted ordinary wave:
$$
a_x=C_x,$$
$$
a_y=\mp iC_x\frac{z_p}{r_L\delta}\eta b_2\sqrt{\frac{\pi}{2}}\exp (i\pi/4).
\eqno (3.9)$$
In case of extraordinary wave (at emission $a_x=0$, $a_y=C_y$), the evolution
of polarization can be treated similarly. The limiting polarization is then
given by
$$
a_x=\mp iC_y\frac{z_p}{r_L\delta}\eta b_2\sqrt{\frac{\pi}{2}}\exp (-i\pi/4)
,$$
$$
a_y=C_y, \eqno (3.10)$$
So the wave mode coupling in the polarization--limiting region causes the
elliptical polarization of the escaping normal waves. The contribution of
circular polarization is characterized by the normalized Stokes parameter
$V$ defined as
$$
V=\frac{i(a_ya_x^\ast -a_xa_y^\ast)}{a_xa_x^\ast +a_ya_y^\ast}. \eqno (3.11)$$
Involving Eqs.\ (3.9), (3.10) one can obtain
$$
\vert V\vert =4\frac{z_p}{r_L\delta}\sin\xi\frac{\vert\alpha -\xi\vert}
{\delta}. \eqno (3.12)$$
It is easy to see that the handedness of the circular polarization 
resulting from the limiting polarization effect remains constant
throughout the pulse.
Note that, to the first order in
$\frac{\textstyle z_p}{\textstyle r_L\delta}$, the polarization profile
appears to be symmetrical, with the peak at the centre. This agrees
qualitatively with the observational data for a number of pulsars (Han
{\it et al.} 1998). 
According to Eq.\ (3.12), the degree of circular polarization is $\sim
\frac{\textstyle z_p}{\textstyle r_L\delta}$. Although this quantity was
regarded as a small parameter, it proved to be not very small at pulsar
conditions, $\frac{\textstyle z_p}{\textstyle r_L\delta}\ge 0.1$ (Cheng
\& Ruderman 1979). Thus, the linearly polarized waves can acquire considerable
circular polarization because of the wave mode coupling in the
polarization-limiting region. In particular, the highest observed circular
polarization, $\sim 60$\%  for PSR 1702-19 (Biggs {\it et.al.} 1988), is
likely to be explained by this effect.

\subsection{The case $\frac{\textstyle z_p}{\textstyle r_L\delta}\gg 1$}
Now we turn to the wave mode coupling in the opposite limit,
$\frac{\textstyle z_p}{\textstyle r_L\delta}\gg 1$. The latter inequality corresponds to
short-period pulsars (Barnard 1986). Substituting Eqs.\ (3.5) and (3.6) into
Eq.\ (2.8) we have, to the first order in $\frac{\textstyle r_L\delta}
{\textstyle z_p}$:
$$
\frac{da_1}{dw}-\frac{i}{4}a_1=\mp\frac{i}{16}\frac{r_L\delta}{z_p\sin\xi}b_2
w^{1/4}a_2-\frac{3i}{8}\frac{r_L\delta}{z_p\sin\xi}b_1w^{1/4}a_1,   $$
$$
\frac{da_2}{dw}=\mp\frac{i}{16}\frac{r_L\delta}{z_p\sin\xi}b_2
w^{1/4}a_1, \eqno (3.13)  $$
with
$$
a_1\equiv b_1a_x-b_2a_y,\quad a_2\equiv b_2a_x+b_1a_y, \quad w\equiv (z_p/z)
^4. \eqno (3.14) $$
Here the polarization-limiting radius, $z_p$, is given by the relation:
$$
\frac{8\omega_p^2(z_p)r_L^2}{\omega c\gamma^3z_p\sin^2\xi}=1. $$
Solving Eq.\ (3.13) to the leading order in $\frac{\textstyle r_L\delta}
{\textstyle z_p}$, one can
trace the evolution of the polarization of the ordinary wave (initially
$a_1=C_1$, $a_2=0 $):
$$
a_1=C_1\exp(iw/4),$$
$$
a_2=\mp C_1\frac{r_L\delta b_2}{16z_p\sin\xi}\left[4w^{1/4}\exp(iw/4)-
\int^ww^{\prime -3/4}\exp(iw^\prime/4)dw^\prime \right]. \eqno (3.15) $$
Setting $w\to 0 $ we find the limiting polarization:
$$
a_1=C_1,$$
$$
a_2=\mp C_1\frac{r_L\delta b_2}{16z_p\sin\xi}\frac{\Gamma(1/4)}{(1/4)^{1/4}}
\exp(i\pi/8). \eqno (3.16)$$
Correspondingly, the limiting polarization of the extraordinary wave
(initially polarized as $a_1=0$, $a_2=C_2$) can be found to be
$$
a_1=\pm C_2\frac{r_L\delta b_2}{16z_p\sin\xi}\frac{\Gamma(1/4)}{(1/4)^{1/4}}
\exp(-i\pi/8),             $$
$$
a_2=C_2. \eqno (3.17)$$
With Eqs.\ (3.16) and (3.17), the normalized Stokes parameter becomes
$$
\vert V\vert =0.2\frac{r_L\delta}{z_p\sin\xi}\frac{\vert\alpha -\xi\vert}
{\delta}. \eqno (3.18)$$
Now the circular polarization is constant along the pulse, with
$\vert V \vert\le 10$\%. Such
values of $V$ are observed for a number of pulsars. Note that in this
case the handedness of circular polarization also remains constant throughout
the pulse.

\section{Conclusion}
We investigated the evolution of normal wave polarization along the trajectory
in the ultrarelativistic highly magnetized electron-positron plasma filling
the magnetosphere of a pulsar. The plasma allows two normal waves which can
ultimately escape from the magnetosphere.
In the emission region, they are linearly
polarized, the radiation emitted being also linearly polarized or
containing an incoherent mixture of radiation with the two linear
polarizations. While propagating inside the magnetosphere, the waves are
influenced by the plasma. So the emergent polarization should be
set up by the propagation effects.

The polarization behaviour was described proceeding from  Maxwell's
equations together with the equation of particle motion in the rotating
magnetosphere. For simplicity we examined the wave mode coupling introduced
by the rotation of a dipolar magnetic field in the limits $\frac{\textstyle
z_p}{\textstyle r_L\delta}\ll 1$ and $\frac{\textstyle z_p}{\textstyle r_L
\delta}\gg 1$, where $z_p$ is the polarization-limiting radius, $r_L$ the
light cylinder radius, $\delta$ the beam width. As a result of this effect,
the emergent radiation should be elliptically polarized, with the sense of
the circular component being the same throughout the pulse. Given that
$r_L\delta\ll z_p<r_L$, the degree of circular polarization is found to be
of a few per cents and to remain constant along the pulse.
In case $z_p\le r_L\delta$ the wave mode coupling can cause considerably
higher circular polarization. In this limit, the polarization profile 
is symmetrical and peaks at the center of the pulse.
Such behaviour is compatible with the observed V-profiles in many pulsars. 
So the observations testify to the case $z_p<r_L\delta$ implying that the 
polarization-limiting radius should lie not too far from the emission region.
This is also supported by the absence of essential decrease in the position 
angle swing as a result of the limiting polarization effect (Lyne \&
Manchester 1988).

\end{document}